\documentclass[12pt]{article}

\usepackage{amssymb}
\usepackage{amsmath}
\usepackage{graphics}
\usepackage{array}
\usepackage{afterpage}
\usepackage{epsfig}
\usepackage{float}

\begin{document}
\title{Nuclear Matter Spectral Functions by Transport Theory\footnote{Work
supported by
BMBF, DFG and GSI Darmstadt}}
\author{J. Lehr~$^{\textrm{a}}$, H. Lenske~$^{\textrm{a}}$,
S. Leupold~$^{\textrm{a}}$ and U. Mosel~$^{\textrm{a,b}}$\\
{\small{$^{\textrm{a}}$~Institut f\"ur Theoretische Physik,
Universit\"at Giessen,}}\\
{\small{Heinrich-Buff-Ring 16, D-35392 Giessen, Germany}}\\
{\small{$^{\textrm{b}}$~Institute for Nuclear Theory,
University of Washington,}}\\
{\small{Box 351550, Seattle, WA 98195, USA}}\\
UGI-01-05} \maketitle
\date{}

\begin{abstract}
Quantum transport theory is used to calculate the nucleon spectral
function in infinite nuclear matter. A self-consistent description
is obtained by utilizing the relations between collision rates and
correlation functions. Static and dynamical self-energies are
taken into account in the single particle propagators. The real
parts of the non-static self-energy contributions are calculated
by dispersion theory thus conserving the analyticity of momentum
distributions. The transport theoretical spectral functions,
momentum distributions, occupation probabilities and responce
functions are in close agreement with results of variational and
other many-body theoretical calculations. The results indicate that the
nucleon spectral functions are determined only by the average short-range
correlation strength.
\end{abstract}


\bigskip

\section{Introduction}

\noindent A still controversial problem of nuclear many-body
theory is the content of short-range correlations in nuclear
matter and finite nuclei. The experimentally observed spectral
functions, e.g. measured in $A(e,e'p)X$ \cite{Witt90} and more
recent $A(e,e'pp)X$ \cite{Rosner98} experiments clearly show the
presence of a sizeable amount of high-momentum processes in nuclei
not accounted for by mean-field dynamics. The observed pattern of
the energy and momentum distributions indicates considerable
admixtures of dynamical short-range correlations beyond the level
of a static mean-field. While typical $(e,e')$ experiments are
performed on stable target nuclei corresponding to densities close
to saturation extreme nucleon removal reactions with dripline
nuclei like $^{11}\textrm{Be}$ and $^{19}\textrm{C}$ \cite{LHK00,RMV01} show
increasing evidence for dynamical correlations also in low density
nuclear matter.

An overall measure of short-range correlations in nuclear matter
is the depletion of ground state momentum distributions from a
pure Fermi gas picture by about 10\%. The deviations are due to
processes scattering nucleons from states inside the Fermi sphere
into high momentum configurations which clearly are not of
mean-field nature. As a result, the momentum distribution obtains
a high momentum tail extending significantly beyond the Fermi
surface. An important finding is that the magnitude and the shape
of the high momentum component is almost independent of the system
under consideration while the low momentum parts, especially in
light nuclei, are affected by the shell structure and finite size
effects. Hence, the high momentum tails of the spectral functions
are likely to reflect a universal property of nuclear many-body
dynamics at short distances.

The whole subject has become of renewed interest by the recent
discussions on off-shell transport theory and in-medium cross
sections \cite{leupold}, structure functions in a nuclear
environment \cite{benhar00} and the QCD-related phenomenon of
color transparency \cite{Strikman,Benhar99}. The standard
approach, taken for granted in most of the theoretical studies, is
to assume that nucleons in a nuclear medium behave similar as in
free space except for a change of the energy-momentum relation
giving rise to a modified single particle spectrum but still
concentrating the strength at the "on-shell" mean-field point. It
has to be realized, however, that such a strict quasi-particle
assumption is only partially supported by the available data. A
general observation, e.g. in $(e,e´p)$ data is that the deviations
increase for states deep inside the Fermi sphere.

Investigations on non-standard phenomena rely on a safe
understanding of the many-body theoretical aspects of short-range
correlations. In fact, the results obtained from many-body theory
describe the available data rather satisfactorily. The majority of
the model calculations for infinite matter are using Brueckner and
Dirac-Brueckner techniques, see e.g.\
\cite{Ramos,FP84,MKP95,DL96,DL97,D98}. An explanation referring to
the QCD aspects of strong interactions was proposed in
\cite{Strikman}. In \cite{Peter} a correlation dynamical treatment
was applied. The Dirac-Brueckner calculations in \cite{DL97},
including hole-hole propagation, led to an extended and
numerically rather involved energy-momentum structure of
self-energies but the net effect on binding energies and
occupation probabilities was surprisingly moderate. Most of the
approaches use the quasi-particle approximation, i.e.\ assuming a
sharp energy distribution for the nucleons in intermediate states
(see e.g.\ \cite{FdJM}); only very recently
self-consistent calculations, going beyond the quasiparticle
approximation, have been performed \cite{dickhoff}.
Occupation probabilities in stable nuclei
could be well described by second RPA \cite{LW90} and by
polarization self-energies \cite{EL89}, respectively, also in
unstable nuclei \cite{RMV01,LHK00}.

In this paper the quantum transport theoretical description of
nucleon spectral functions in nuclear matter presented in
\cite{lehr00} is discussed in more detail and extended in several
aspects by considering additional observables. Since a long time
transport theory is known to be {\em in principle} the correct
description \cite{KB62,BM90,DB} for spectral functions. But
explicit calculations on a realistic level were pending, also
because an appropriate treatment of off-shell effects in transport
equations was missing. The recent progress on implementing
off-shell transport for heavy-ion and other nuclear
collisions numerically \cite{ef_off,ef_dilep,cj} and theoretically
\cite{leupold} were important pre-requisites for the studies in
\cite{lehr00} and the present work.

The theoretical background formulae are summarized in section
\ref{theory}. Since collision rates and correlation functions are
directly related, the calculation of either of the two quantities
depends on the knowledge of the other one. Theoretically, this
rather involved self-consistency problem cannot be solved in
closed form, but a practical approach is obtained by an iterative
sequence of successive approximations for self-energies, spectral
functions and collision integrals \cite{DL97,lehr00}.

Details of the numerical realization of the many-body theoretical
relations
are discussed in section \ref{numerics}. In extension of the
previous work \cite{lehr00} here we include also dispersive
self-energies into the single particle propagators. These
non-static, i.e. energy and momentum dependent, self-energies are
calculated by dispersion theory. In this way, the consistency of
single particle dynamics and spectral functions is enforced and,
equally important, the analyticity of the results is guaranteed.
The equations are solved using an average matrix element
accounting primarily for the "hard" short-range collisions.
Different to \cite{lehr00} here we include an energy and momentum
dependent form factor.
In section \ref{numerics} results for nuclear
matter spectral functions, ground state momentum distributions,
occupation probabilities and response functions are presented. The
agreement with results obtained with many-body theory by Benhar et
al. \cite{benhar} and Ciofi degli Atti et al. \cite{C90,C95} is
striking. It confirms the universality of short-range correlations
in nuclear systems and leads us to the conclusion that they are
determined by an average matrix element representing globally the
hard in-medium collisions.


\section{Spectral Functions in Quantum Transport Theory}\label{theory}

\subsection{Transport Theoretical Relations}\label{transport}

Here, the known fundamental relations of quantum transport theory
\cite{KB62,BM90} for the description of non-stationary processes
in an interacting quantum system are briefly summarized, following
closely the presentation in \cite{lehr00}. In quantum transport
theory the non-stationary processes which introduce a coupling
between causal and anti-causal single particle propagation are
described by the one-particle correlation functions
\begin{eqnarray}\label{corrf}
g^>(1,1')&=&-i\langle \Psi(1)\Psi^\dagger(1')\rangle \nonumber \\
g^<(1,1')&=&i\langle \Psi^\dagger(1')\Psi(1)\rangle,
\end{eqnarray}
where $\Psi$ are the nucleon field operators in Heisenberg
representation. Correspondingly, in an interacting quantum system
the single particle self-energy operator includes correlation
self-energies $\Sigma^{<>}$ which couple particle and hole degrees
of freedom \cite{KB62,BM90}. Clearly, $g^{<>}$ and $\Sigma^{<>}$
are closely related. The wanted relation is obtained from
transport theory. After a Fourier transformation to
energy-momentum representation the self-energies are found as
\cite{KB62}
\begin{eqnarray}
  \label{eq:sigma>}
\Sigma^>(\omega,p)&=&g\int{d^3p_2d\omega_2\over(2\pi)^4}
{d^3p_3d\omega_3\over
(2\pi)^4}{d^3p_4d\omega_4\over(2\pi)^4}(2\pi)^4\delta^4(p+p_2-p_3-p_4)
\, \overline{\vert{\cal{M}}\vert^2}   \nonumber \\
    & &{} \times
    g^<(\omega_2,p_2)g^>(\omega_3,p_3)g^>(\omega_4,p_4),
\end{eqnarray}
\begin{eqnarray}
  \label{eq:sigma<}
\Sigma^<(\omega,p)&=&g\int{d^3p_2d\omega_2\over(2\pi)^4}
{d^3p_3d\omega_3\over
(2\pi)^4}{d^3p_4d\omega_4\over(2\pi)^4}(2\pi)^4\delta^4(p+p_2-p_3-p_4)
\, \overline{\vert{\cal{M}}\vert^2} \nonumber \\
    & &{} \times
    g^>(\omega_2,p_2)g^<(\omega_3,p_3)g^<(\omega_4,p_4).
\end{eqnarray}
Here, $g=4$ is the spin-isospin degeneracy factor and
$\overline{\vert{\cal{M}} \vert^2}$ denotes the square of the
in-medium nucleon-nucleon scattering amplitude, averaged over spin
and isospin of the incoming nucleons and summed over spin and
isospin of the outgoing nucleons.

Since both $g^{<>}$ and $\Sigma^{<>}$ describe the correlation dynamics,
the spectral
function can be obtained from either of the two quantities as the
difference over the cut along the energy real axis.
In terms of the correlation propagators, the spectral density is
defined by
\begin{equation}\label{spf}
a(\omega,p)=i\left( g^>(\omega,p)-g^<(\omega,p) \right).
\end{equation}
Non-relativistically, the single particle spectral function is explicitly
found as
\begin{equation}
 \label{eq:spectral}
a(\omega,p)={\Gamma(\omega,p)\over(\omega-{p^2\over 2 m_N}-
\textrm{Re}\Sigma(\omega,p))^2+\frac{1}{4}\Gamma^2(\omega,p)},
\end{equation}
including the particle and hole nucleon self-energy
$\Sigma$.
The width $\Gamma$ is given by the imaginary
part of the retarded self-energy,
\begin{equation}
  \label{eq:gamma}
  \Gamma(\omega,p)=2\textrm{Im}\Sigma(\omega,p)=i(\Sigma^>(\omega,p)-
\Sigma^<(\omega,p)).
\end{equation}
In the limiting case of vanishing correlations, i.e. $\textrm{Im}\Sigma\to 0$,
the usual delta-like quasi-particle spectral function is recovered.

The correlation functions $g^{<>}$ can be re-written in terms of the
phase-space distribution function $f(\omega,p)$:
\begin{align}
  \label{eq:g<>}
    g^<(\omega,p) &= i a(\omega,p)f(\omega,p), \\
    g^>(\omega,p) &= -i a(\omega,p)(1-f(\omega,p)).
\end{align}
For a system at $T=0$, $f$ reduces to
\begin{equation}
f(\omega,p)=\Theta(\omega_F-\omega)
\end{equation}
with the Fermi energy $\omega_F$.
As a result,we obtain for the self-energies the conditions
\begin{align}
&\Sigma^>(\omega,p)=0,\quad \Gamma(\omega,p)=-i\Sigma^<(\omega,p)\quad
\textrm{for}\quad \omega\le \omega_F \\
&\Sigma^<(\omega,p)=0,\quad \Gamma(\omega,p)=i\Sigma^>(\omega,p)\quad
\textrm{for}\quad \omega\ge \omega_F.
\end{align}

\begin{figure}[H]
  \begin{center}
\epsfig{file=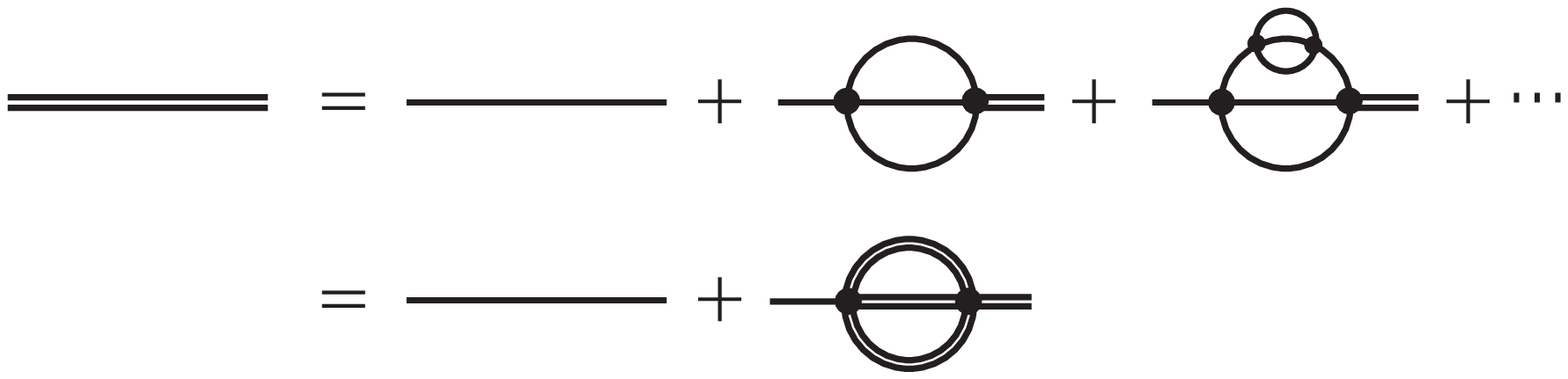,width=14cm}
    \caption{Diagrammatic structure of the self-consistent single particle propagator, indicated by
    a full line. The Dyson equation, corresponding to the complete resummation of the "sun-set" diagrams
    to all orders, is depicted in the second line.}
\label{fig:diagram}
  \end{center}
\end{figure}

Due to the dependence of the width and the single particle spectral function
upon each other
the calculation of $a(\omega,p)$ requires a self-consistent treatment.
Therefore, the transport theoretical approach leads to single particle
propagators including correlation self-energies to all orders by a
complete resummation of the basic {\it sunset} diagram. The
diagrammatic structure of Dyson equation defining the correlated
propagators is shown in Fig. \ref{fig:diagram}.

\subsection{The Scattering Amplitude}\label{me}

Obviously, the correlation effects depend on the yet undetermined
matrix element $\overline{{\cal{M}}}$. The question arises which
processes contribute most significantly to $\overline{{\cal{M}}}$.
Since the bulk of interactions giving rise to long-range
mean-field interactions is already taken care of by the proper
self-energies of particle and hole states the correlation
functions and self-energies must both be determined by those parts
of the fundamental interactions producing non-stationary effects
beyond the mean-field. An estimate of the relevant interaction
scales can be obtained by considering the physical situation,
implying strong contributions from the high-momentum components of
wave functions and interactions. From the observation that bulk
self-energies include in their exchange parts momenta up to twice
the Fermi momentum $k_F$ a realistic estimate is to identify
processes involving momentum transfers $q \gg 2k_F$ as the origin
for short-range correlations. At the saturation point of nuclear
matter with $k_F \sim 270$ MeV/c this corresponds to collisions
with $q \gg 600$ MeV/c. Such processes are located in the
interaction regime of the $\omega$ and $\rho$ vector mesons,
typically accounting for the short-range repulsion in
nucleon-nucleon interactions. These processes are appropriately
described by the short-range parts of a Brueckner G-matrix,
accounting for the whole series of repeated meson-exchange
processes.

In addition to the ladder-type interactions from the exchange of
individual mesons also three-body (and possibly even higher order)
interactions will contribute to short-range correlations in a
nuclear medium. Such interactions involve the intermediate
excitation of nucleon resonances which subsequently decay back to
the nucleon sector by an interaction with a third nucleon. Of
similar importance are virtual excitations of $N\overline{N}$
pairs giving rise to the so-called $Z$ graph contributions
\cite{DL98,Coon}. The major contributions of these sub-nucleonic
processes are known to be expressible to a good approximation in
terms of an effective density dependent two-body interaction (see
e.g. \cite{Urbana}), and, as such, will be part of
$\overline{{\cal{M}}}$. The importance of such processes is
indicated e.g. by the variational nuclear matter calculations of
the Urbana group \cite{Urbana} showing a sizeable contribution
from three-body interactions by subsequent pion exchange already
at and below saturation density. Interestingly, Dirac-Brueckner
calculations including the polarization of the nucleonic Dirac sea
\cite{DL98} and investigations of three-nucleon interactions
\cite{Coon} also point to the importance of dynamically generated
repulsive short-range modes.

In this context it is of interest that the off-shell behavior of
meson-exchange interactions is essentially determined by the mass
of the exchanged boson. Hence, the matrix element
$\overline{{\cal{M}}}$ will vary only weakly on off-shell momenta,
at least on a scale of about $1\textrm{GeV}/c$, allowing to replace it to a
good approximation by an energy and momentum independent constant,
corresponding to an effective contact interaction. A similar
approach is used in Landau-Migdal theory \cite{Migdal} describing
interactions by the in-medium $NN$ forward scattering amplitude at
the Fermi surface. Different to conventional Landau-Migdal theory
and to Ref.~\cite{lehr00}, here we will account for the remaining
dependences at large momentum transfers by a global off-shell form
factor $F(\omega, \vec{q})$, to be discussed below.

Taken together, these arguments lead to the expectation that
$\overline{{\cal{M}}}$ is a quantity being almost independent of
the specific system and the momentum transfer. The matrix element
will reflect general properties of short-range nucleonic many-body
dynamics, mediated by {\em hard} processes (on the scale of
typical nuclear momenta) of universal character. Sensitivity to a
single, specific channel seems to be unlikely considering the
experience that short-range correlations are typically determined
by a collection of interfering processes with strong mutual
cancellations (see e.g. \cite{Coon}). We assume universality and
treat $\overline{{\cal{M}}}$ as a global parameter.


\section{Numerical Approach and Results}\label{numerics}

\subsection{Details of the Calculation}

Diagrammatically, the particle- and hole-type transition rates
$\Sigma^>$ and $\Sigma^<$, Eqs. (\ref{eq:sigma>}) and
(\ref{eq:sigma<}), are of two-particle--one-hole (2p1h) and
one-particle--two-hole (1p2h) structure, respectively, as shown in
Fig. \ref{fig:diagram}. In this respect, they are of the same basic
structure as the polarization self-energies considered in
many-body theoretical descriptions. However, while in many-body
theory the polarization self-energies are typically included
perturbatively in lowest order only by performing the integrations
over intermediate 2p1h and 1p2h states with quasi-particle
spectral functions, e.g. in Ref.~\cite{benhar} and also Refs.
\cite{LW90,EL89}, a more extended scheme accounting for higher
order effects was presented in \cite{lehr00}, where
the correlation self-energies with the self-consistently obtained spectral
functions were calculated. This allows a
non-perturbative summation of the whole series of $n$p $m$h
intermediate states as shown in \cite{lehr00}.

However, here we
take advantage of the observation that already the first iteration
agrees within a few percent with the final result of a fully
iterated calculation. Because the main effect of higher order
iterations is to redistribute a minor fraction of the strength
into the high momentum tails of spectral functions the results
presented here remain accurate on the energy-momentum scale
relevant for the gross structures of spectral function and
momentum distribution in pure nucleonic matter. Hence, these
results also can be considered as re-confirming the validity of
the perturbative treatment, assumed implicitly in previous
investigations.

As already pointed out in \cite{lehr00} the numerical
simplifications achieved when neglecting the off-shell dependence
of $\cal{M}$ are at the expense of violating analyticity, seen
very clearly in an unrealistic behavior of the momentum
distribution for $p \sim p_F$. These problems are corrected by
introducing an energy and momentum dependent form factor

\begin{equation}\label{eq:formf}
F(\omega_{\textrm{tot}},\vec p_{\textrm{tot}})=
{\Lambda^4\over \Lambda^4+\Bigl(\omega_{\textrm{tot}}
+{\vec p_{\textrm{tot}}^2\over 4 m_N}\Bigr)^4}
\end{equation}
that multiplies the average matrix element. Here
$\omega_{\textrm{tot}}=\omega+\omega_2=\omega_3+\omega_4$, $\vec
p_{\textrm{tot}}=\vec p+\vec p_2=\vec p_3+\vec p_4$ and
$\Lambda=m_N$. The structure of the form factor was adopted from
those applied in $K$ matrix calculations (see e.g.
\cite{feuster}). Note that the form factor is symmetric under the
exchange of the incoming and outgoing nucleon pair as seen from
the delta function in Eqs. (\ref{eq:sigma>}), (\ref{eq:sigma<}).
It is apparent that collisions at total energies and momenta
beyond the scale of the cut-off $\Lambda$ will be suppressed by
the form factor. Because of the fourth power structure and with
our choice $\Lambda= m_N$ processes with on-shell momenta of
up to $\sim 3k_F$ are not affected by the form factor, especially
leaving collisions inside the Fermi sphere untouched.

Theoretically, the main advantage of using a form factor is that
the polarization contribution $\Sigma_D$ to the real part of the
single particle self-energies can now be calculated explicitly and
consistently by means of a subtracted dispersion relation
\begin{equation}\label{eq:dispersion}
\Sigma_D(\omega,p)=({p^2\over 2m_N}-\omega)\
{\cal{P}}\intop_{-\infty}^\infty {d\omega^\prime\over
2\pi}{\Gamma(\omega^\prime,p)\over(\omega-\omega^\prime)
({p^2\over2 m_N}-\omega^\prime)},
\end{equation}
where the on-shell point $\omega={p^2\over 2m_N}$ is chosen as
subtraction point. Note, that according to the above definition
$\Sigma_D(\omega,p)$ is a real quantity.

In order to account for the static mean-field a density dependent
but energy and momentum independent real self-energy $\Sigma_0$ is
added, i.e. the real part of the self-energy becomes in total
Re$\Sigma(q,\omega)=\Sigma_0+\Sigma_D(q,\omega)$.
Eq.~\ref{eq:dispersion} corresponds to assume that dispersive
on-shell self-energies are already included in the ''mean-field''
$\Sigma_0$, following closely the approach of phenomenological
mean-field theory. In fact, the constant $\Sigma_0$ only serves to
define the scale for the excitation energy $\omega$ which in our
case is given by $\omega\geq \omega_F$. Since we are not
interested in the dependence of our results on the density but
only study the system at nuclear saturation density, we can absorb
$\Sigma_0$ into $\omega$, i.e. we re-define $\omega$ by
$\omega+\Sigma_0$ for the following. Compared to \cite{lehr00},
this corresponds to a shift of the energy scale by about 52.6 MeV.

In general, $\Sigma_0$ will also include momentum dependent
contributions, reflecting the non-localities of the Hartree-Fock
mean-field and the on-shell part of $\Sigma_D$, respectively. An
estimate in the effective mass approximation, i.e. absorbing
3-momentum dependent effects into the kinetic energy term, showed
that this type of momentum dependence primarily leads to a
rescaling of energies by a (density dependent) constant given by
the ratio of the effective mass to the bare mass. Conservation of
probability requires to multiply a corresponding factor to the
spectral functions. The net result is that spectral functions are
almost unchanged if both scaling effects are taken properly into 
account. Because we are primarily interested in investigating the
energy-momentum structure of dispersive off-shell self-energies we
neglect such contributions, also for the sake of having a minimal
set of free parameters.

Starting from the vacuum spectral function of the nucleon
\begin{equation}
  \label{eq:vac_spectral}
  a_{\textrm{initial}}(\omega,p)=2\pi\delta(\omega-{p^2\over2 m_N}),
\end{equation}
one obtains from Eqs.(\ref{eq:sigma>}), (\ref{eq:sigma<}) the
following expression for the self-energies:
\begin{equation}
  \label{eq:imsigma_delta}
  \begin{split}
\Sigma^{\overset{<}{>}}(\omega,p) &= {\pm i g\over 4\pi^3}
  \intop_0^\infty dk \intop_0^\infty dq{kq^2 m_N\over p}\
\overline{\vert{\cal{M}}_0\vert^2}\ F^2\
\Theta(2m_N\omega+{k^2\over2}-2q^2+p^2+2pk)\\
 & \times\Theta(2pk-2m_N\omega-{k^2\over2}+2q^2-p^2)\\
 &\times\Theta(\pm{k^2\over2}\pm 2q^2\mp 2m_N\omega\mp p_F^2)\
 \Theta(\pm p_F^2\mp {k^2\over4}\mp q^2) \\
&\times\Bigl( \Theta(\pm p_F-q\mp {k\over2})\pm
  {p_F^2- {k^2\over 4}- q^2\over kq}\
\Theta(q\pm {k\over 2}\mp p_F)\Bigr).
\end{split}
\end{equation}
Here $\vec k=\vec p_3+\vec p_4$, $\vec q=(\vec p_3-\vec p_4)/2$ and
$p_F^2=2m_N\omega_F$. The upper sign refers to $\Sigma^<$, the
lower one to $\Sigma^>$.

The average $NN$ (off-shell) scattering amplitude ${\cal{M}}_0$, which
determines the self-energies and spectral functions, was treated
as a universal parameter. Adjusting it to the spectral functions
from many-body calculations of Benhar et al. \cite{benhar} we
derive $\left(\overline{\vert{\cal{M}}_0\vert^2}\right)^{1/2}$=207
MeV fm$^{3}$.

Relating the derived
$\left(\overline{\vert{\cal{M}}_0\vert^2}\right)^{1/2}$ to
on-shell processes would correspond to a constant total $NN$ cross
section of about 20~mb accounting for roughly $2/3$ of the
commonly used value. The above strength, however, compares very
well with the in-medium scattering amplitude of
$+221.0~\textrm{MeV\ fm}^3$ derived by Landau-Migdal theory from
the nuclear equation of state for the Urbana interaction model
\cite{Urbana}, averaged over the Fermi sphere and spin and
isospin. Since the calculations of \cite{benhar} were performed in
the correlated basis function approach, including also the
three-body nucleon interactions in the Urbana prescription
\cite{Urbana}, we consider the derived value for
$\left(\overline{\vert{\cal{M}}_0\vert^2}\right)^{1/2}$ as a
realistic overall measure for correlations in nuclear matter.

\subsection{Nucleon Self-Energies and Spectral Functions}

\begin{figure}[H]
  \begin{center}
\epsfig{file=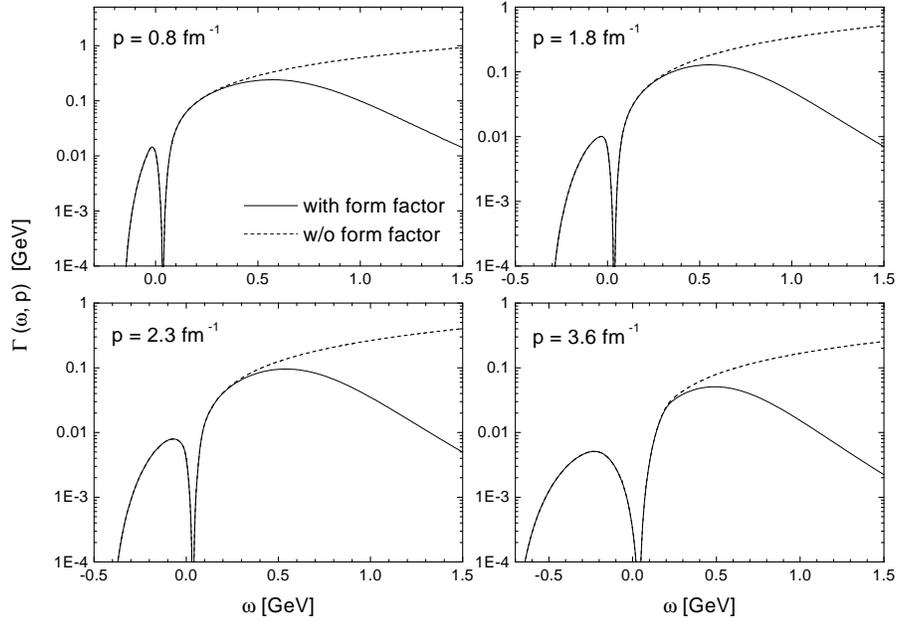,width=14cm}
    \caption{Widths of the nucleon spectral distributions
    for different momenta, calculated with (full) and without
    (dashed) form factor. Mainly the high-energy parts are affected by
    the off-shell form factor.}
\label{fig:gamma_nucl}
  \end{center}
\end{figure}

Before discussing the full spectral function
Eq.~(\ref{eq:spectral}), it is worthwhile to consider the influence
of the form factor on the width and the spectral function without
the dispersive $\Sigma_D$ contribution. In
Fig.~\ref{fig:gamma_nucl} we show the results for the spectral
nucleon width using i) the constant matrix element described above
and ii) the additional form factor from Eq.~(\ref{eq:formf})
for different momenta. It is worth noting that
the influence of the form factor is completely different for
energies below and above the Fermi energy: while for
$\omega>\omega_F$ there is a strong suppression of the width for
energies larger than 0.5 GeV, for $\omega<\omega_F$ the results
for the calculations with and without form factor almost coincide.
The reason for the weak influence for $\omega<\omega_F$ can be
found in the step functions in Eq.~(\ref{eq:imsigma_delta}). The
nucleons are treated on-shell, thus Eq.~(\ref{eq:formf}) reduces to
\begin{equation*}
    F(k,q)={\Lambda^4\over \Lambda^4+\Bigl({k^2\over 2m_N}+{q^2\over m_N}
\Bigr)^4}.
\end{equation*}
Hence, for hole-type states where $k^2/2+q^2<2p_F^2$ (see fourth
step function in Eq.~(\ref{eq:imsigma_delta}) for $\Sigma^<$),
$F^2>0.999$ deviates only insignificantly from unity.

\begin{figure}[H]
  \begin{center}
\epsfig{file=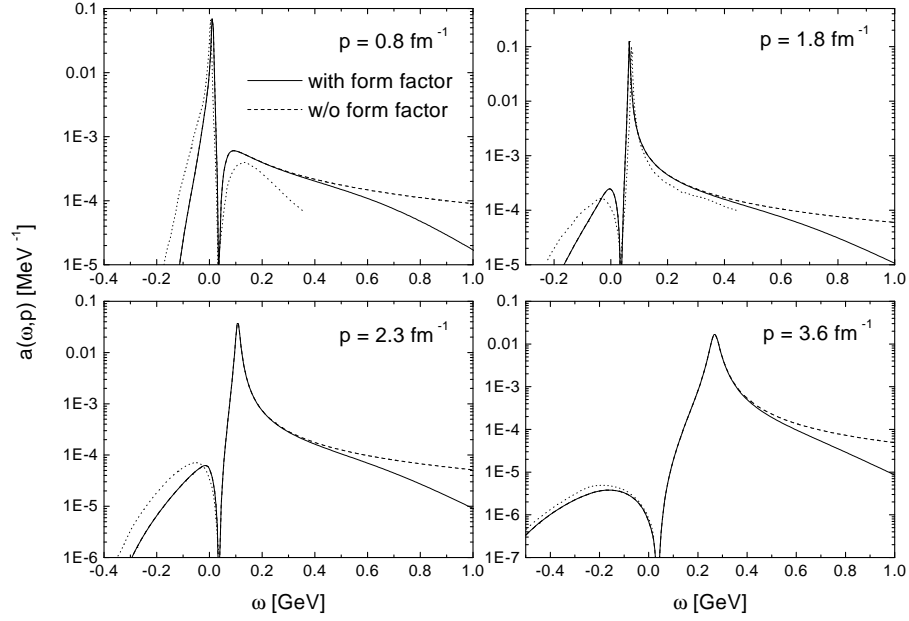,width=14cm}
    \caption{Influence of the form factor on the spectral function.
    Dispersive self-energies are not included. The dotted lines show
    the results from Benhar et al. \protect\cite{benhar}.}
\label{fig:spec_ff}
  \end{center}
\end{figure}

The observed behavior of the width in the calculations with and
without form factor transmits to the spectral function, as can be
seen in Fig.~\ref{fig:spec_ff}. The form factor strongly affects
the high energy tails of the particle-type spectral functions at
$\omega > \omega_F$ while for energies below $\omega_F$ the
changes are negligible. This is a very satisfying result showing
that the momentum distribution
\begin{equation}
  \label{eq:mom_distr}
  n(p)=\intop_{-\infty}^{\omega_F}{d\omega\over 2\pi}a(\omega,p)
\end{equation}
is almost independent of the form factor.
Also shown in Fig.~\ref{fig:spec_ff} are the results from Benhar et al.
\cite{benhar}.

\begin{figure}[H]
  \begin{center}
\epsfig{file=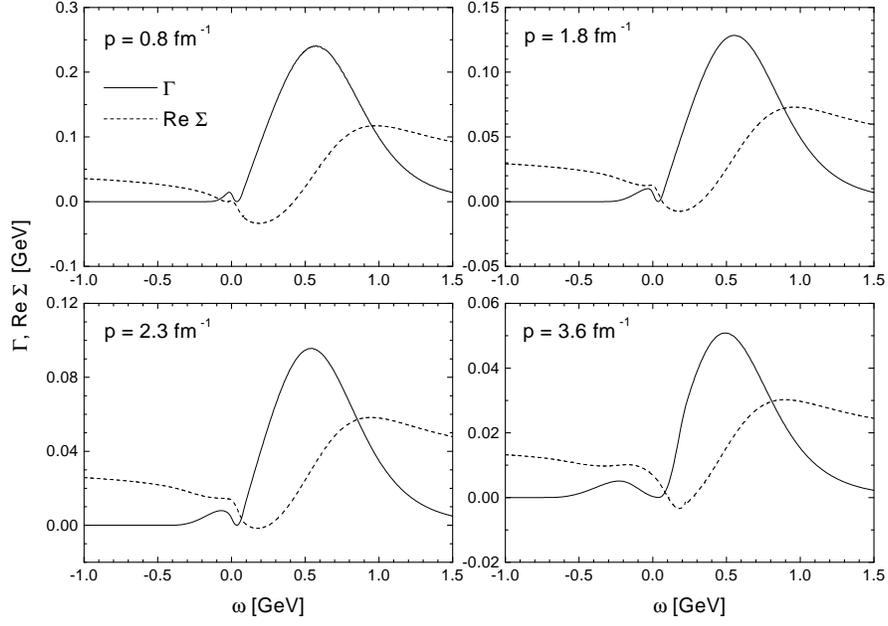,width=14cm}
    \caption{Width of the nucleon spectral function (full line) and real part
of the dispersive self-energy (dotted line) for different
momentum cuts.} \label{fig:real_im}
  \end{center}
\end{figure}

\begin{figure}[H]
  \begin{center}
\epsfig{file=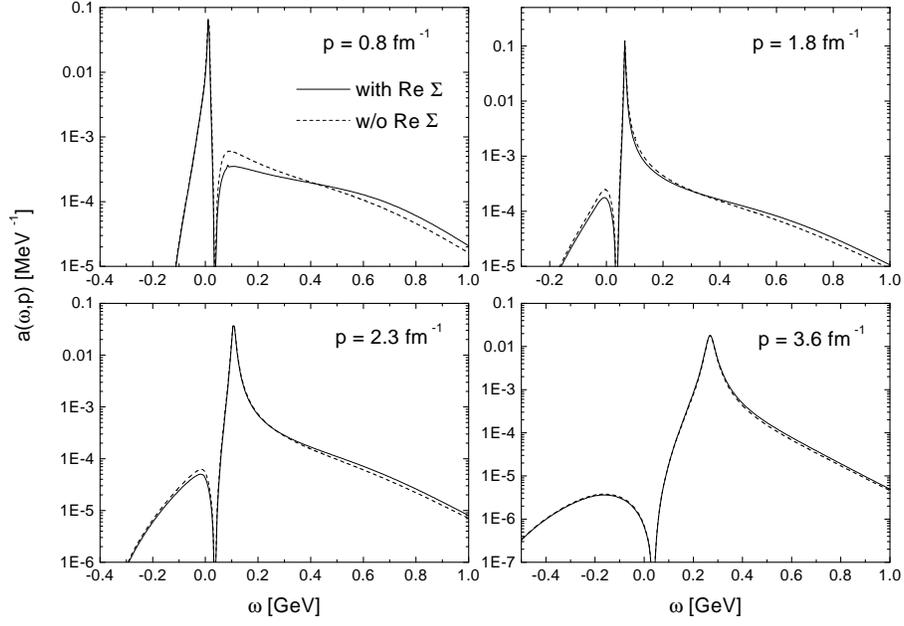,width=14cm}
    \caption{Nucleon spectral function at momenta below and above the Fermi
    momentum.
    Results with (full line) and without (dotted line)
    the real part of the dispersive self-energy are displayed. The
    quasi-particle peaks are clearly visible.}
    \label{fig:spec_res}
  \end{center}
\end{figure}

We are now in the position to calculate the full spectral
function, including also the dispersive self-energy $\Sigma_D$,
Eq.~(\ref{eq:dispersion}). In Fig.~\ref{fig:real_im} results for
$\Sigma_D$ and $\Gamma=2\textrm{Im}\Sigma$ are displayed. Due to
the decrease of $\Gamma$ with increasing momentum also $\Sigma_D$
is reduced. Therefore, the dispersive real part is expected to
vanish at high $p$. This can be seen in Fig.~\ref{fig:spec_res},
where the results for the spectral function with and without
$\Sigma_D$ are presented.  At the on-shell point the difference of
the two curves vanishes because $\Sigma_D({p^2\over
2m_N},p)\equiv 0$ by definition.
Comparing to Fig.~\ref{fig:spec_ff} the combined action of the form
factor and $\Sigma_D$ gives rise to a slight re-distribution of
the strengths in the tail regions but leaves the region around the
quasi-particle peak almost unaffected.

\begin{figure}[H]
  \begin{center}
\epsfig{file=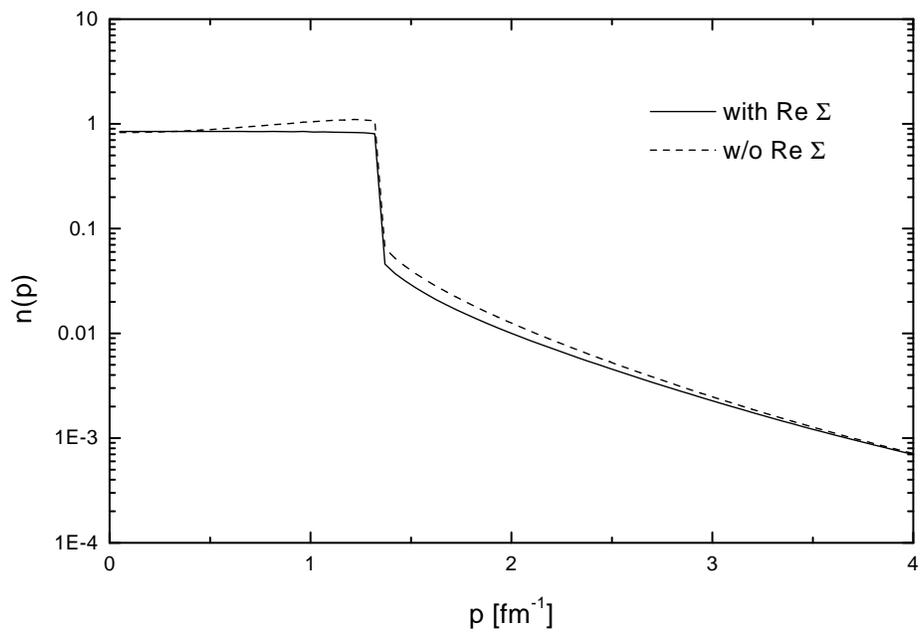,width=14cm}
    \caption{Nucleon momentum distribution in nuclear matter.
Results neglecting analyticity (dashed) are also displayed.}
\label{fig:mom_distr}
  \end{center}
\end{figure}

The influence of the analyticity of the self-energies on the
momentum distribution (\ref{eq:mom_distr}) is illustrated in
Fig.~\ref{fig:mom_distr}. Indeed, as anticipated before, the
behavior of $n(p)$ below $p_F$ is totally changed from a strong
increase to a smooth decrease towards $p_F$ being now in agreement
with many-body results e.g. \cite{DL97,benhar}. For momenta above $p_F$ the
differences between the two approaches are almost negligible.

\subsection{Occupation Probabilities}

Analyticity directly affects the normalization of the spectral
function,
\begin{equation}
  \label{eq:norm}
  N(p)=\intop_{-\infty}^{\infty}{d\omega\over 2\pi}a(\omega,p),
\end{equation}
where conservation of probability requires $N(p)\equiv 1$. Whereas
the calculation without $\Sigma_D$ violates this
condition - varying with momentum and reaching values of up to 30
\% - the inclusion of the dispersive self-energy
$\Sigma_D$ leads to a properly normalized spectral
function over the full momentum range.

\begin{figure}[H]
  \begin{center}
\epsfig{file=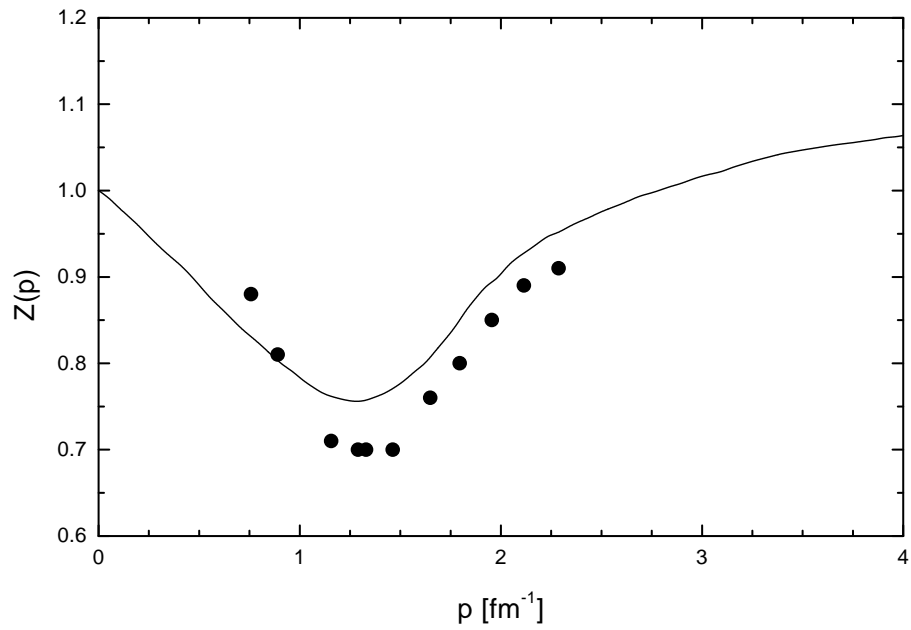,width=14cm}
    \caption{Nucleon spectroscopic factor in nuclear matter
    evaluated for the on-shell energy (see Eq.~(\ref{zfac})).
    For comparison, results from the full many-body calculation of
    Benhar et al. \protect\cite{benhar} (dots) are also shown.}
\label{fig:zfac}
  \end{center}
\end{figure}

The physical significance of $\Sigma_D$ becomes evident by
considering
\begin{equation}\label{zfac}
Z(\omega,p)=\frac{1}{1-{\partial \Sigma_D(\omega,p)\over\partial
\omega}}
\end{equation}
which at the pole positions is known to describe the energy (and
momentum) dependent wave function renormalization or spectroscopic
factor accounting for the dissipation of single particle strength
into the more complex many-body configurations. Results for
$Z(\omega,p)$ at the on-shell energy $\omega=\frac{p^2}{2m}$ are
displayed in Fig.~\ref{fig:zfac}. The overall shape follows
roughly the spectroscopic factors from the many-body calculations
of Benhar et al. \cite{benhar}, also shown in the figure for
comparison. In magnitude our results are different, being higher by
about 7\% at the Fermi surface.

\subsection{The Density Response of Nuclear Matter}

Inclusive scattering of high-energetic electrons off nuclei
measures directly the longitudinal and transversal density
response functions from which information on the density-density
correlation function is obtained \cite{Czyz63}. Moreover, the
nuclear response function is of current interest because it enters
directly into investigations of scaling behavior and color
transparency \cite{benhar00,Benhar99}. At large momenta the
density response is expected to be determined by universal
properties of nuclear systems, being independent of the system
under consideration.

\begin{figure}[H]
  \begin{center}
\epsfig{file=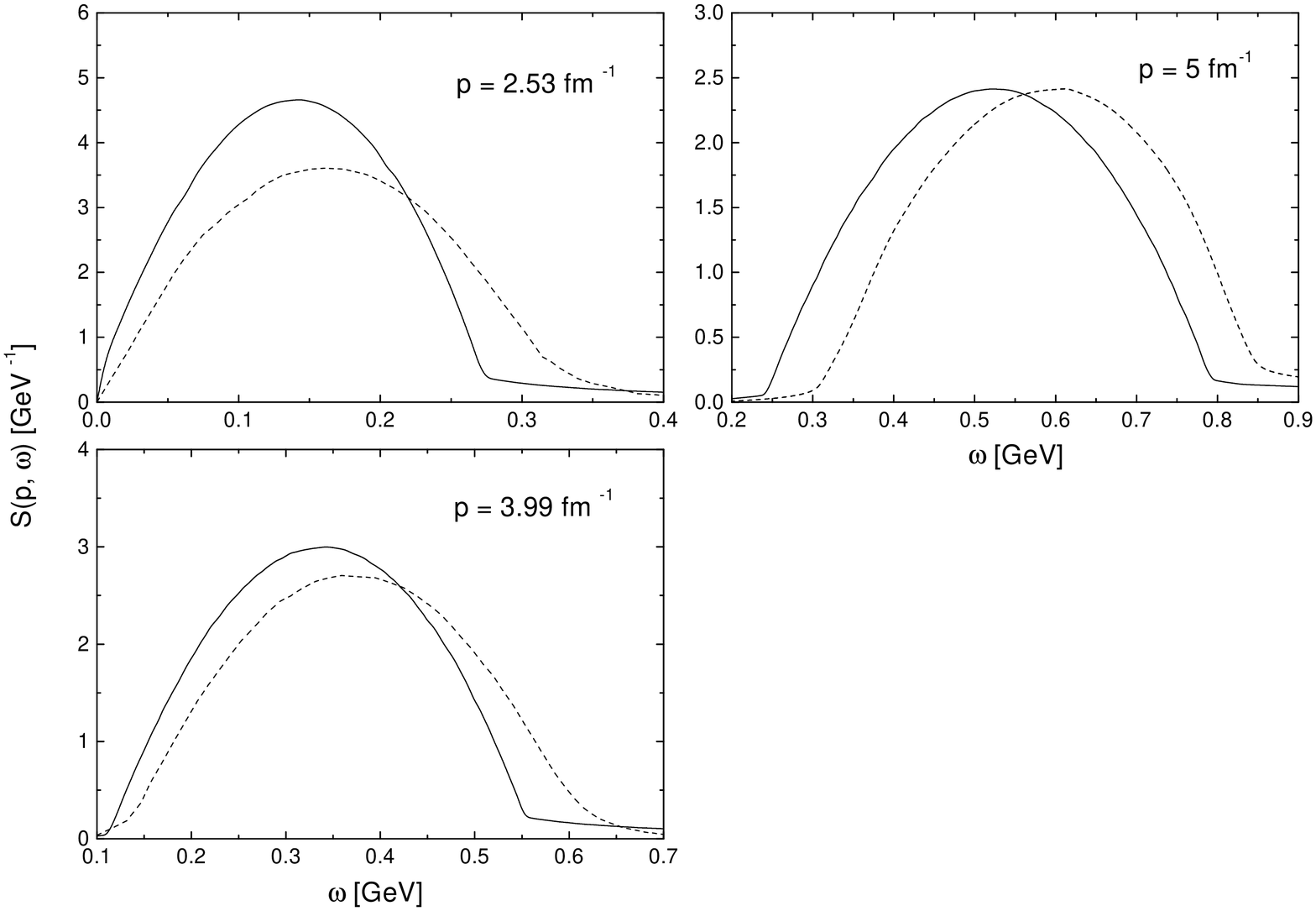,width=14cm}
    \caption{Nuclear matter density response functions, Eq.~(\ref{response}),
at different momenta. The dashed lines show the results from
\protect\cite{benhar,benhar01}.}
\label{fig:response}
  \end{center}
\end{figure}

Here, we neglect final state interactions (FSI) of the struck
nucleon with the bulk and consider the density response of
infinite nuclear matter in plane wave impulse approximation (PWIA)
only. In the non-relativistic limit the density response (per
energy) is defined by
\begin{equation}\label{response}
S(\omega,p)=\frac{1}{V_F}\int{ {d\varepsilon\over
2\pi}{\frac{d^3k}{(2\pi)^3} a(\varepsilon,k)
\delta\biggl(\omega+\varepsilon-{\vert\vec p+\vec k\vert^2\over
2m_N}\biggr) \Theta(\vert\vec p+\vec k\vert-k_F)}},
\end{equation}
where, according to our definition of spectral functions a
normalization to the volume of the Fermi sphere,
\begin{equation}
V_F=\int{\frac{d^3k}{(2\pi)^3}\Theta(k_F-k)},
\end{equation}
was introduced. The spectral function $a(\varepsilon,k)$ is seen
to act in Eq.~(\ref{response}) as a source for the energy and
momentum distribution of the ejected nucleons. At large momentum
transfers where the FSI become negligibly small the PWIA results
are expected to carry already the relevant information. In
Fig.~\ref{fig:response} the PWIA nuclear matter response function
for several momenta are shown.  In each case the maximum appears
at the quasi-elastic (on-shell) energy
$\omega_{QE}=\frac{p^2}{2m}$, as typical for quasi-free
scattering. The widths of the distributions correlates with the
Fermi momentum. Results of Benhar et al. \cite{benhar,benhar01}
are also given in this figure as dashed lines.

Considering the strong simplifications of the present approach the
overall agreement in shape and magnitude is remarkable. A common
feature, however, is a persistent shift of our response functions
to lower energies. From Eq.~\ref{response} is found that
$S(\omega,p)$ is a non-trivial cut through the energy-momentum
structure of the spectral function, varying rapidly with the
external energy and momentum. The deviations seen
Fig.~\ref{fig:spec_ff} for the spectral functions become more
enhanced in the response functions. By optimizing the form factor
the agreement both for the spectral and response functions,
respectively, could be improved. For the present purpose, however,
this was not done, because an optimal fit to other results is not
of primary interest.

\section{Summary and Conclusions}\label{conclusions}

Correlations and nucleon spectral functions in nuclear matter were
described by transport theory. Dispersive contributions to the
nucleon self-energies are included, thus guaranteeing analyticity
of spectral functions and momentum distributions. An important
pre-requisite for this achievement was to include a form factor.
The comparatively hard cut-off $\Lambda=m_N$ agrees with the
assumed structure of the average matrix element
$\overline{\cal{M}}$ as mainly given by processes involving
momentum transfers well beyond twice the Fermi momentum of the
system. The insensitivity of high momentum processes on bulk
properties of a nuclear system and their weak off-shell dependence
led us to describe dynamics by a momentum independent, universal
matrix element, treated as the only adjustable parameter of the
approach, once the form factor has been fixed. Spectral functions
and momentum distributions for infinite nuclear matter at
equilibrium were presented.

The energy and momentum dependences of spectral functions and
momentum distributions are well reproduced by our calculations,
confirming also the conjectures of Danielewicz and Bertsch
\cite{DB}. Considering the strongly simplified approach the
agreement with results from many-body calculations is impressive
and supports the approach.

The results lead to the conclusion that the hole-type spectral
functions and momentum distributions are dominated by phase space
effects rather than by the off-shell momentum structure of
interactions. A residual dependence on off-shell effects, however,
is observed, indicating that the approach is most reliable for
processes with intermediate energy and momentum transfers up to
1~GeV. Still, these effects could be accounted for globally by an
overall form factor of a rather simple functional form. In the
hole sector the main effect of the form factor is to ensure
analyticity, leaving the shape of the spectral functions almost
unaffected. In the particle region, of course, the spectral
function depends more strongly on off-shell effects.
Interestingly, the different behaviour in the particle and hole
regions closely reflects again phase space effects because the
density of 2p-1h configurations - responsible for the damping of
particle states - increases much more rapid than the number of
2h-1p states entering into the hole self-energies. Hence, the
particle spectral functions reacts much more sensitive to the
cut-off properties of interactions at high off-shell momenta.

The shape of the momentum distribution is related to the magnitude
of $\overline{\cal{M}}$: Increasing the value - corresponding to a
stronger interaction amongst the nucleons - would increase the
occupation of states above $p_F$ and soften the Fermi edge. This
close relationship constrains the admissible range of values. To
the extent that higher order self-consistency is of minor
importance, as indicated by our calculations, this also allows, in
principle, to extract $\overline{\cal{M}}$ from spectral functions
and the slopes of momentum distributions by precise measurements
of the high momentum tails. However, we again emphasize that such
a determination will only provide average information on
short-range correlations in nuclei, not giving access to specific
processes. Moreover, from the observed independence of $n(p)$ on
form factor effects we expect that also the off-shell properties
of interactions will play a minor role for momentum distributions,
except for global properties as analyticity. This is easily
understood if we accept that {\em hard} processes mediated by the
(single or multiple) exchange of heavy mesons are the major
source. Since the off-shell behavior of meson exchange
interactions is essentially fixed by the mass of the meson it is
clear that the off-shell momentum dependence of the underlying
interactions will become important only on momentum scales well
beyond 1 GeV.

\vspace{1cm}


\newpage

\begin{thebibliography}{99}
\bibitem{Witt90} for a review see, e.g., P.K.A. de Witt Huberts, J. Phys. G
                 (Nucl. Part. Phys.) {\bf 16} (1990) 507.
\bibitem{Rosner98}G. Rosner, in {\em Perspectives in Hadronic Physics}
                  edited by S. Boffi, C. Ciofi degli Atti and M.M.
                  Giannini (World Scientific, Singapore, 1998), p. 185.
\bibitem{LHK00}H. Lenske, F. Hofmann, C.M. Keil,
               Rep. Prog. Nucl. Part. Phys. {\bf 46} (2001) 187.
\bibitem{RMV01}D. Cortina-Lopez {\em et al.}, Euro. Phys. J. {\bf A10}
            (2001) 49.
\bibitem{leupold} S. Leupold, Nucl. Phys. {\bf A672} (2000) 475,
nucl-th/0008036, Nucl. Phys. {\bf A}, in print.
\bibitem{benhar00}O. Benhar, V.R. Pandharipande, I. Sick, Phys. Lett. {\bf B489}
 (2000) 131.
\bibitem{Strikman}L. Frankfurt and M.I. Strikman, Phys. Rep. {\bf 76} (1981) 215.
\bibitem{Benhar99}O. Benhar, Phys. Rev. Lett. 83 (1999) 3130.
\bibitem{Ramos}A. Ramos, A. Polls and W.H. Dickhoff, Nucl. Phys. {\bf A503} (1990) 1;
A. Ramos, A. Polls and W.H. Dickhoff, Phys. Rev. {\bf C43} (1991)
2239.
\bibitem{FP84}S. Fantoni and V.R. Pandharipande, Nucl. Phys. {\bf A427} (1984) 473.
\bibitem{MKP95}H. M\"uther, G. Knehr and A. Polls, Phys. Rev. {\bf C52} (1995) 2955.
\bibitem{DL96}F. de Jong and H. Lenske, Phys. Rev. {\bf C54} (1996) 1488.
\bibitem{DL97}F. de Jong and H. Lenske, Phys. Rev. {\bf C56} (1997) 154.
\bibitem{D98}W.H. Dickhoff, Phys. Rev. {\bf C58} (1998) 2807.
\bibitem{Peter}A. Peter, W. Cassing, J.M. H\"auser and A. Pfitzner,
Nucl. Phys. {\bf A573} (1994) 93.
\bibitem{FdJM} F. de Jong and R. Malfliet, Phys. Rev. {\bf C44} (1991)
998.
\bibitem{dickhoff}  W.H. Dickhoff, talk at the Workshop on Correlations in
Nucleons and Nuclei, INT, University of Washington, March 2001.

\bibitem{LW90}H. Lenske and J. Wambach, Phys. Lett. {\bf B249} (1990) 377.
\bibitem{EL89}F.J. Eckle, H. Lenske, G. Eckle, G. Graw, R. Hertenberger,
H. Kader, F. Merz, H. Nann, P. Schiemenz and H.H. Wolter, Phys.
Rev. {\bf C39} (1989) 1662; F.J. Eckle, H. Lenske, G. Eckle, G.
Graw, R. Hertenberger, H. Kader, H.J. Maier, F. Merz, H. Nann, P.
Schiemenz and H.H. Wolter, Nucl. Phys. {\bf A506} (1990) 159.


\bibitem{lehr00}J. Lehr, M. Effenberger, H. Lenske, S. Leupold, U.
Mosel, Phys. Lett. {\bf B483} (2000) 324.
\bibitem{KB62}L.P. Kadanoff and G. Baym, Quantum Statistical Mechanics
(Benjamin, New York, 1962).
\bibitem{BM90} W. Botermans and R. Malfliet, Phys. Rep. {\bf 198} (1990) 115.
\bibitem{DB}G. Bertsch and P. Danielewicz, Phys. Lett. {\bf B367} (1996) 55.
\bibitem{ef_off}M. Effenberger and U. Mosel, Phys. Rev. {\bf C60} (1999)
051901.
\bibitem{ef_dilep}M. Effenberger, E.L. Bratkovskaya and U. Mosel, Phys. Rev.
{\bf C60} (1999) 044614.
\bibitem{cj} W. Cassing and S. Juchem, Nucl. Phys. {\bf A665} (2000) 377;
Nucl. Phys. {\bf A672} (2000) 417; Nucl. Phys. {\bf A677} (2000) 445.

\bibitem{benhar} O. Benhar, A. Fabrocini and S. Fantoni, Nucl. Phys.
{\bf A505} (1989) 267; O. Benhar, A. Fabrocini and S. Fantoni,
Nucl. Phys. {\bf A550} (1992) 201.
\bibitem{C90}C. Ciofi degli Atti, S. Simula, L.L. Frankfurt and M.I.
Strikman, Phys. Rev. {\bf C44} R7 (1991).
\bibitem{C95}C. Ciofi degli Atti and S. Simula, Phys. Rev. {\bf C53} 1689
(1996).
\bibitem{DL98}F. de Jong and H. Lenske, Phys. Rev. {\bf C58} (1998) 890.
\bibitem{Coon}S.A. Coon, M.T. Pe$\tilde{\textrm{n}}$a, Phys. Rev. {\bf C48}
(1993) 2559;
S.A. Coon, M.T. Pe$\tilde{\textrm{n}}$a, D.O. Riska, Phys. Rev. {\bf C52}
(1995) 2925.
\bibitem{Urbana} A. Akmal, V.R. Pandharipande, D.G. Ravenhall, Phys. Rev.
{\bf C58} (1998) 1804.
\bibitem{Migdal}A.B. Migdal, E.E. Saperstein, M.A. Troisky, D.N. Voskresensky,
Phys. Rep. {\bf 192} (1990) 179.
\bibitem{feuster} T. Feuster and U. Mosel, Phys. Rev. {\bf C58} (1997) 457.


\bibitem{Czyz63}W. Czyz, K. Gottfried, Ann. Phys. {\bf 45} (1963)
47.
\bibitem{benhar01} O. Benhar, A.Fabrocini and S.Fantoni,
Phys. Rev. Lett. {\bf 87} (2001), 052501.


\end{thebibliography}
\end{document}